\documentclass[useAMS,usenatbib]{mn2e} 
\usepackage{aas_macros}
\usepackage{graphics}
\usepackage{epsfig}  
\usepackage{natbib} 
\usepackage{float}
\newcommand{\Vec}[1]{{\bf #1}}
\newcommand{\Tab}[1]{Table~\ref{#1}}
\newcommand{\Sec}[1]{Section~\ref{#1}}
\newcommand{\Eq}[1]{Eq.(\ref{#1})}
\newcommand{\Fig}[1]{Fig.\ref{#1}}
\newcommand{\hMpc}{{\ifmmode{h^{-1}{\rm Mpc}}\else{$h^{-1}$Mpc}\fi}}
\newcommand{\hkpc}{{\ifmmode{h^{-1}{\rm kpc}}\else{$h^{-1}$kpc}\fi}}
\newcommand{\hMsun}{{\ifmmode{h^{-1}{\rm {M_{\odot}}}}\else{$h^{-1}{\rm{M_{\odot}}}$}\fi}}

\def\lesssim{\mathrel{\hbox{\rlap{\hbox{\lower4pt\hbox{$\sim$}}}\hbox{$<$}}}}
\def\gtrsim{\mathrel{\hbox{\rlap{\hbox{\lower4pt\hbox{$\sim$}}}\hbox{$>$}}}}

\title[Radial alignment in the obersver's plane]
      {The radial alignment of dark matter subhalos: from simulations to observations}
\author[Knebe et al.] 
{Alexander Knebe$^1$, Hideki Yahagi$^{2,3,4,5,6}$, Hiroyuki Kase$^{2,7}$, Geraint Lewis$^8$, 
\newauthor Brad K. Gibson$^{5,8}$
  \\
  $^1$Astrophysikalisches Institut Potsdam, An der Sternwarte 16, Germany
  \\
  $^2$Department of Astronomy, University of Tokyo, Tokyo 113-0033, Japan
  \\
  $^3$Division of Theoretical Astronomy, National Astronomical Observatory, Japan
  \\
  $^4$Research Fellow of the Japan Society for the Promotion of Science
  \\
  $^5$Centre for Astrophysics, University of Central Lancashire, Preston PR1 2HE, UK
  \\
  $^6$Research Institute for Information Technology, University of Kyushu, Fukuoka 812-8581, Japan
  \\
  $^7$present address: Prometech Inc., 7-3-1 Hongo, Bunkyo, Tokyo 113-0033, Japan
  \\
  $^8$School of Physics, University of Sydney, Sydney NSW 2006, Australia
  \\
  }

\begin{document}

\date{Submitted Version ...}

\pagerange{\pageref{firstpage}--\pageref{lastpage}} \pubyear{2008}

\maketitle

\label{firstpage}

\begin{abstract}
  We explore the radial alignment of subhalos in 2-dimensional
  projections of cosmological simulations. While most other recent
  studies focussed on quantifying the signal utilizing the full
  3-dimensional spatial information any comparison to observational
  data has to be done in projection along random lines-of-sight. We
  have a suite of well resolved host dark matter halos at our disposal
  ranging from $6\times 10^{14}$\hMsun\ down to $6\times
  10^{13}$\hMsun. For these host systems we do observe that the major
  axis of the projected 2D mass distribution of subhalos aligns with
  its (projected) distance vector to the host's centre. The signal is
  actually stronger than the observed alignment. However, when
  considering only the innermost 10-20\% of the subhalo's particles
  for the 2D shape measurement we recover the observed correlation. We
  further acknowledge that this signal is independent of subhalo mass.
\end{abstract}


\section{Introduction}
\label{sec:introduction}

The concordance of a multitude of recent cosmological studies has
demonstrated that we appear to live in a spatially flat,
$\Lambda$-dominated cold dark matter ($\Lambda$CDM) universe
\cite[cf. ][]{Spergel07}. During the past decade simulation codes and
computer hardware have advanced to such a stage where it has been
possible to resolve in detail the formation of dark matter halos and
their subhalo populations in a cosmological context
\citep[e.g. ][]{Klypin99}. These results, coupled with the
simultaneous increase in observational data (e.g. 2 degree Field
galaxy redshift survey (2dFGRS), \citet{2dF-DR}; Sloan Digital Sky
Survey (SDSS), \citet{SDSS-DR5}), have opened up a whole new window on
the concordance cosmogony in the field that has become known as
``near-field cosmology'' \citep{Freeman02}.

One particular property of the satellite population that has caught
the attention of simulators recently is the radial alignment of their
primary axes of subhalos with respect to the distance vector to the
host's centre. The first evidence for this effect was reported for the
Coma cluster, where it was observed that the projected major axes of
cluster members preferentially align with the direction to the cluster
centre \citep{Hawley75, Thompson76}.  Such a correlation between
satellite elongation and radius vector has further been confirmed by
statistical analysis of the SDSS data \citep{Pereira05, Agustsson06,
  Faltenbacher07a, Wang07}. The radial alignment of subhalo shapes
towards the centre of their host has also been measured for the
subhalo population in cosmological simulations \citep{Knebe08,
  Kuhlen07, Faltenbacher07b, Pereira07}. We though note that all these
authors used the 3-dimensional spatial information available to
them. However, any fair comparison to the signal found observationally
requires the projection of the simulation data into an observer's
plane, i.e. averaging over a substantial number of 2D restrictions of
the data \citep[cf. ][]{Faltenbacher07b}. This is the primary
motivation for the present study: \textit{Can we still find a signal
  of radial alignment when limiting the simulation data to 2D?}

In this \textit{Letter} we provide evidence that the radial alignment
of subhalos in cosmological simulations -- when projected into two
dimensions -- is substantially stronger than found in observations. We
though recover the observed correlation strength when restricting the
shape measurement to the innermost 10-20\% of the subhalos' particles;
this result supports the findings of \cite{Faltenbacher07b} who also
used the inner regions of their satellite halos as a proxy for the
orientation of a hypothetical galaxy and found good agreement with the
measurements from SDSS data. We further show that the signal does not
depend on the mass of the actual subhalo.

\section{Method}
\label{sec:method}

\subsection{Simulations}
\label{sec:simulations}
The cosmological $N$-body simulation employed in this study was
generated as part of the Numerical Galaxy Catalog program
\citep{Nagashima05}.  Utilizing a parallel version of an established
$N$-body code \citep{Yahagi01, Yahagi05}, the algorithm uses an
adaptive mesh refinement technique to maintain accuracy
\citep[cf. ][]{Kravtsov97, Knebe01, Yahagi01, Teyssier02}.  The
simulation considered a region of the universe with a boxsize of 70
$h^{-1}$ Mpc (comoving), with $512^3$ particles, resulting in a
particle mass of $3.04 \times 10^8$\hMsun.  The adopted cosmological
parameters represent a concordance $\Lambda$CDM model with
$\Omega_m=0.3$, $\Omega_\Lambda=0.7$, $h=0.7$, and $\sigma_8=0.9$
\citep[cf. ][]{Spergel03}. The simulation was started at $z=41$ and
analysed at $z=0$.

\subsection{(Sub-)Halo Identification}
\label{sec:subhalos}
In extracting groups and subhalos, the standard friends-of-friends
(FoF) method was employed \citep{Davis85}, with the ratio of the
linking length to the mean separation of particles $b=0.2$, and
cluster and group scale host halos were choosen to have masses is in
the range, $6.45 \times 10^{13}$\hMsun $\leq M_{\rm host} \leq 5.95
\times 10^{14}$\hMsun.  Once these host halos were identified, a
hierarchical finder based on the FoF (HFoF) method \citep{Klypin99}
used to isolate subhalos, employing s successively decreasing linking
length.  Furthermore, the evaporative method was used to discard
unbound subhalos \citep{Pfitzer97}, and all unbound particles were
removed from the subhalos. At this stage, we calculated potential
energy of particles in subhalos iteratively, defining a minimum
subhalo as containing 100 particles.  \footnote{We note that this is a
  conservative criterion as other studies indicate that halo
  catalogues are complete for objects containing in excess of $\geq$50
  particles \citep[e.g.][]{Kravtsov04b, Allgood06}.}  While this is
very time consuming procedure, the process was accelerated by
employing a special purpose computer for self-gravitating system,
GRAPE-6 \citep{Makino03}; details of the subhalo finding are given in
\citet{Kase07}.

We ended up with 40 host halos with a combined number of 2188 subhalos
in excess of 100 particles. In the following analysis we stack the
information from all these hosts and present the results in one single
plot. We are confident not to obscure any signal as it has recently
been shown that the radial alignment is independent on the host mass
\citep{Knebe08}.

When investigating the dependence of radial alignment on the mass of
the subhalos we split the subhalos into three mass bins,
i.e. $[10^{11},10^{12}]$\hMsun, $[10^{12},10^{13}]$\hMsun, and
$[10^{13},10^{14}]$\hMsun. The total number of subhalos in each of
these bins (as well as the number of subhalos with $M<10^{11}$\hMsun\
is given in \Tab{tab:subhalos}. We further like to note 
that the most massive subhalo weights $M_{\rm sub}^{\rm max} = 2.9
\times 10^{13}$\hMsun\ and belongs to a $M_{\rm host} = 1.6 \times
10^{14}$\hMsun\ host halo.

\begin{table}
\begin{center}
  \caption{Number of subhalos in a certain mass range.}
\begin{tabular}{cc}
\hline
$M$ & $N_{\rm sub}$ \\
\hline
$M < 10^{11}$\hMsun              & 1199  \\
$M\in [10^{11}, 10^{12}]$ \hMsun &  846  \\
$M\in [10^{12}, 10^{13}]$ \hMsun &  120  \\
$M\in [10^{13}, 10^{14}]$ \hMsun &   23  \\
\end{tabular}
\label{tab:subhalos}
\end{center}
\end{table}

\section{Results}
\label{sec:results}
Our view of clusters of galaxies upon the sky is a projection of the
true, three dimensional distribution and orientation of halos and
their associated galaxies. In this study, we therefore consider
projected views of our simulated halos, adopting 19 projections in
both azimuthal angle and cosine of latitudinal angle, respectively.
The results presented subsequently in this study average over these
differing orientations.

\subsection{Subhalo Shapes in 2D}
\label{sec:shapes}

The principal axes of the projected two-dimensional density
distributions of all our subhalos were calculated using the reduced moment of inertia
tensor \citep{Katz91, Dubinski91}\footnote{We like to note in passing that it makes little difference using either the reduced or the "standard" moment of inertia tensor that does not include the $1/\zeta_l^2$ term.},
\begin{eqnarray}\label{eqn:zeta}
\hat{I}_{i,j} &=& \frac{1}{N_p} \sum_{l=1}^{N_p}
\frac{x_{l,i} x_{l,j}}{\zeta_l^2},\\
\zeta_l^2 &=& x_l^2 + \left(\frac{y_l}{s}\right)^2,
\end{eqnarray}


\noindent
where $x_{l,i}$ is $i$-th component of projected (2-dimensional)
position of the $l$-th constituent particle, and $N_p$ is the number
of particles in the subhalo.  Solving for the eigensystem of this
tensor identifies the principle axes of the distribution.  We then
define the sphericity
\begin{equation}\label{eqn:sphericity}
s=\frac{a_2}{a_1}
\end{equation}

\noindent
of a (projected) subhalo as the ratio of the minor to major axis. The
procedure in determining these principle axes has to be done
iteratively, starting with the assumption of sphericity $s=1$. The
eigensystem is solved for $s$ until the value converges.

\subsubsection{Mass Dependence}
\label{sec:s2Dmass}
\begin{figure}
\begin{center}
\begin{minipage}{0.40\textwidth}
        \epsfig{file=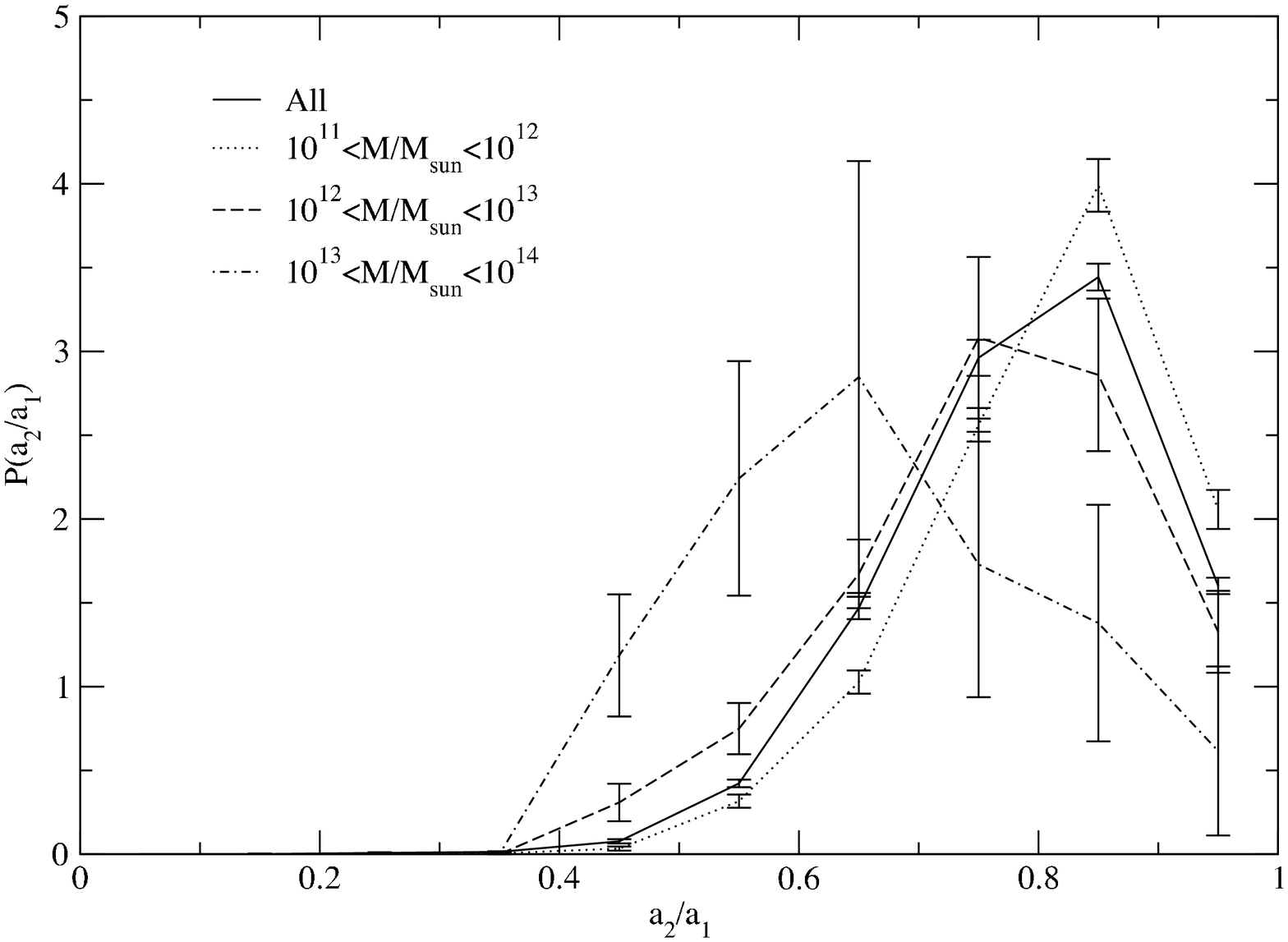, width=1\textwidth, angle=0}
\end{minipage}
\end{center}
\caption{Distribution of the 2D sphericity $s=a_2/a_1$. Solid line indicates the distribution of axis ratio using all
subhalos. The different lines represents restrictions of the subhalo masses under consideration to the intervalls $[10^{11},10^{12}]$\hMsun, $[10^{12},10^{13}]$\hMsun, and $[10^{13},10^{14}]$\hMsun. Error bars indicate the square root of the
sum of the Poisson error and the projection variation.
 \label{fig:s2Dmass}}
\end{figure}

In \Fig{fig:s2Dmass} we examine the variation of the (normalized)
probability distribution $P(s=a_2/a_1)$ of sphericity with subhalo
mass. We observe a (marginal) trend for $s$ to decrease with
increasing subhalo mass -- as reported by \citet{Allgood06} for field
halos. We also note that the projected sphericities are of similar
value to the 3D sphericities as found in \citet{Knebe08} making a
study of radial alignment feasible.  We need to stress that the number
statistics for the subhalo mass bin $[10^{13}, 10^{14}]$\hMsun\ is
rather small (cf. \Tab{tab:subhalos}); there are a mere 23 subhalos in
the respective mass range and hence the large error bars. As these
objects are in fact rather massive and comparable to the actual host
halos, we are not surprised to find sphericities closer to those of
isolated/field halos \citep[$\langle s\rangle \approx 0.66$,
cf. ][]{Frenk88, Warren92, Kasun05, Bailin05, Allgood06, Maccio07,
  Bett07}

\subsubsection{Radial Dependence}
\label{sec:s2Drad}
\begin{figure}
\begin{center}
\begin{minipage}{0.40\textwidth}
        \epsfig{file=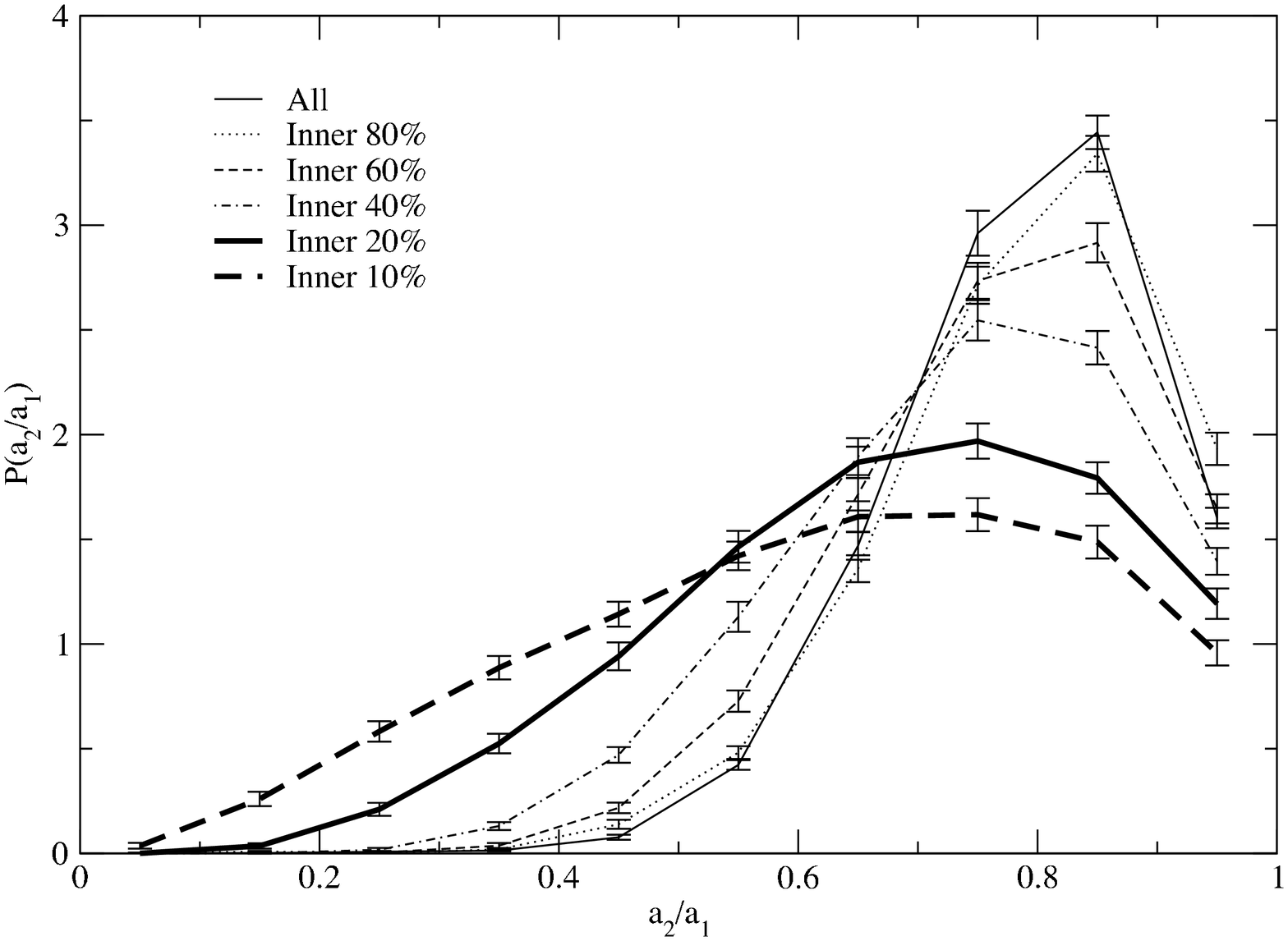, width=1\textwidth, angle=0}
\end{minipage}
\end{center}
\caption{Distribution of the 2D sphericity $s=a_2/a_1$. 
The different lines represent shape measures at different
mass thresholds. Error bars indicate the projection variation.
 \label{fig:s2Drad}}
\end{figure}

As with many cosmological $N$-body studies, we are considering the
orientations of \textit{dark matter} (sub-)halos and are comparing
them with the distributions of \textit{baryons} observed in galaxy
clusters. Such a comparison can be significantly biased if the
observed orientation of galaxies, which occupy the deepest part of the
subhalo potential, are not aligned with the global orientation of the
dark matter \citep[cf. ][]{Bailin05}. Hence we also present the
variation of subhalo sphericity (as well as the radial alignment in
\Sec{sec:ra2Drad}) with distance to its centre.

To select the inner parts of each subhalos, $\zeta_l$ was calculated
assuming $s=0$.  Particles were then sorted by $\zeta_l$ and the inner
subhalo was selected as a numerical fraction of the subhalo's mass.
Once this distribution was defined, the moment of inertia tensor was
recalculated, and the eigensystem solved iteratively to determine the
inner value of $s$ (as outlined in \Sec{sec:shapes}).

\Fig{fig:s2Drad} shows that the inner regions of subhalos appear to be
significantly more elliptical than their overall matter
distribution. This is in agreement with the findings of others who
also reported that dark matter halos tend to increase their
asphericity towards the centre \citep[e.g. ][]{Jing02, Allgood06,
  Hayashi07}. We will return to this finding in the following Section
when investigating the radial alignment as a function of distance to
the centre: Is the signal actually enhanced due to the increase in
sphericity? Or does the correlation weaken despite this result?

We tough like to note that \citet{Allgood06} also cautioned that the
sphericity may be overestimated if too few particles are used. In
order to check the credibility of the trend seen in \Fig{fig:s2Drad} we
recalculated the distributions restricting the haloes to have masses
in the range $M\in [10^{12}, 10^{13}]$\hMsun. We still observe the
same shift in the peak of the distributions; however, the extended tail
towards lower $s$-values for the inner 10-20\% has
vanished, i.e. there are hardly any objects with $s<0.4$.

\subsection{Radial Alignment in 2D}
\label{sec:ra2D}
The principal aim of this study is to investigate whether or not there
is a dependence of the radial alignment of subhalos (i.e. the
alignment of their major axis with respect to the centre of the host)
on the mass of their host halo when observing the mass distributions
in projection. Recent observational evidence suggests that the major
axis (in projection) of satellite galaxies tend to ``point towards the
centre of their host'' \citep[e.g. ][]{Pereira05, Agustsson06,
  Faltenbacher07a}. It is therefore natural to ask whether or not
subhalos in cosmological simulations display a similar trend. To date
a few studies have investigated this subject \citep{Knebe08, Kuhlen07,
  Faltenbacher07b, Pereira07} all of which used the full 3D spatial
information available to them; only \citet{Faltenbacher07b} also
presented a 2D analysis. However, a fair comparison to observational
findings requires a projection of the simulation data into the
observer's plane.

To measure the radial alignment of our (projected) subhalos, we use
the eigenvector $\Vec{E}_{a_1}$ which corresponds to the direction of
the major axis $a_1$ of the subhalo. We quantify the radial alignment
of subhalos as the angle between the major axis $\Vec{E}_{a_1}$ of
each subhalo and the radius vector of the subhalo in the reference
frame of the host:
\begin{equation}\label{eqn:ra}
\cos \phi = \frac{\Vec{R}_{\rm sub} \cdot \Vec{E}_{a_1, \rm sub}}{|\Vec{R}_{\rm sub}| |\Vec{E}_{a_1, \rm sub}|} \ .
\end{equation}

\subsubsection{Mass Dependence}
\label{sec:ra2Dmass}

\begin{figure}
\begin{center}
\begin{minipage}{0.40\textwidth}
        \epsfig{file=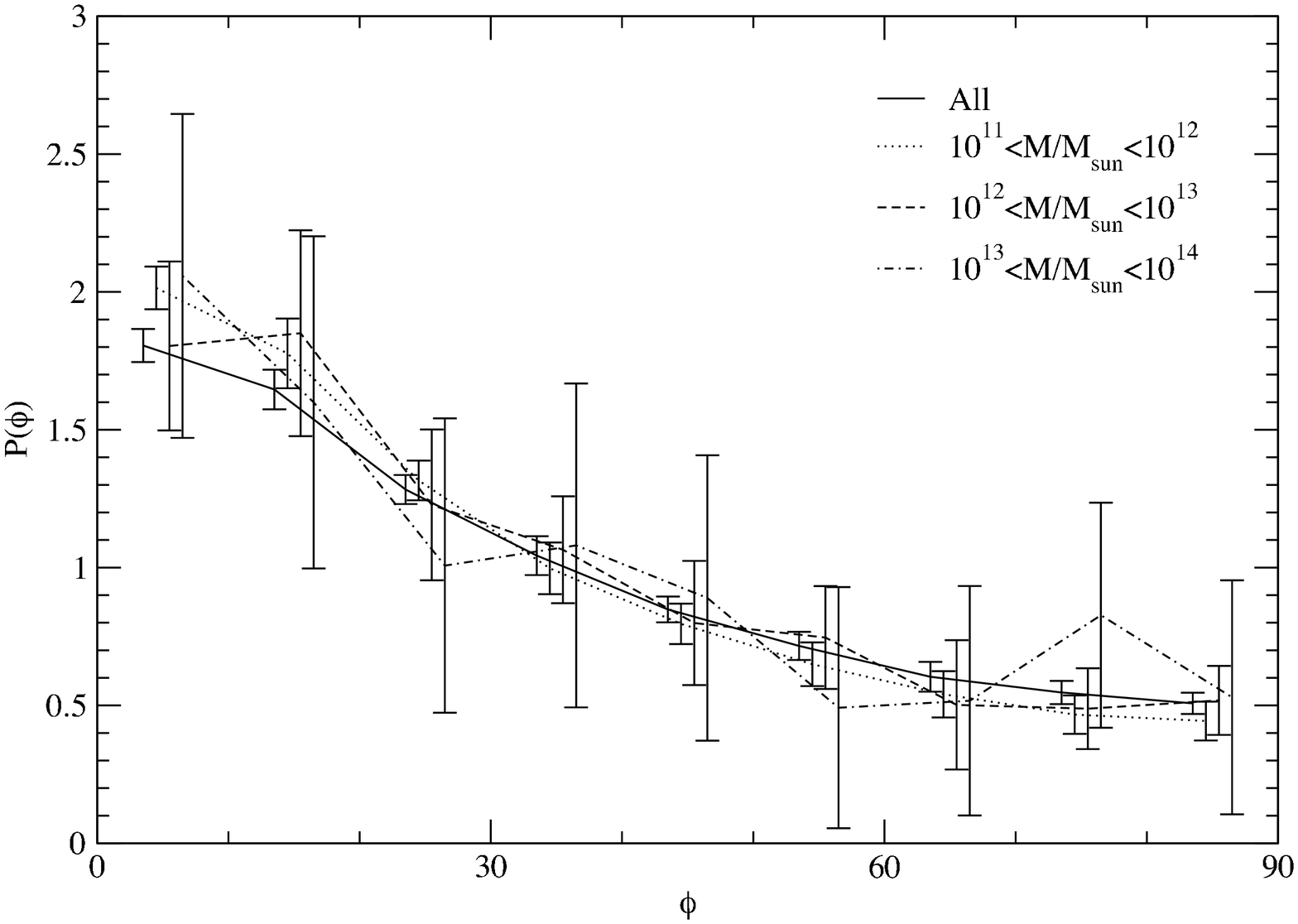, width=1\textwidth, angle=0}
\end{minipage}
\end{center}
\caption{Correlation of angle between major axes of subhalos and
position of the subhalos from the center of their host.  We use the same subhalo mass bins as in \Fig{fig:s2Dmass}. Error bars (marginally shifted for clarity) are again the square root of the Poisson error and the projection variation.
\label{fig:ra2Dmass}}
\end{figure}

In \Fig{fig:ra2Dmass} we show the normalized distribution of the angle
$\phi$ as defined via \Eq{eqn:ra}. While the black line shows the
signal for all subhalos under consideration, the different lines are
representative for various mass bins. We find a positive radial
alignment signal different from isotropy, in agreement with other
studies \citep{Knebe08, Kuhlen07, Faltenbacher07b, Pereira07} -- even
when restricting the analysis to 2D.  We further conclude from this
plot that no particular mass range is responsible for the signal. We
also note that the strength of the signal is substatially stronger
than the one seen in observations \citep[cf. ][]{Pereira05}. We will
though see in the following Section that this sensitively depends on
the point where we measure the actual (projected) shape of the
subhalo.

\subsubsection{Radial Dependence}
\label{sec:ra2Drad}
\begin{figure}
\begin{center}
\begin{minipage}{0.40\textwidth}
        \epsfig{file=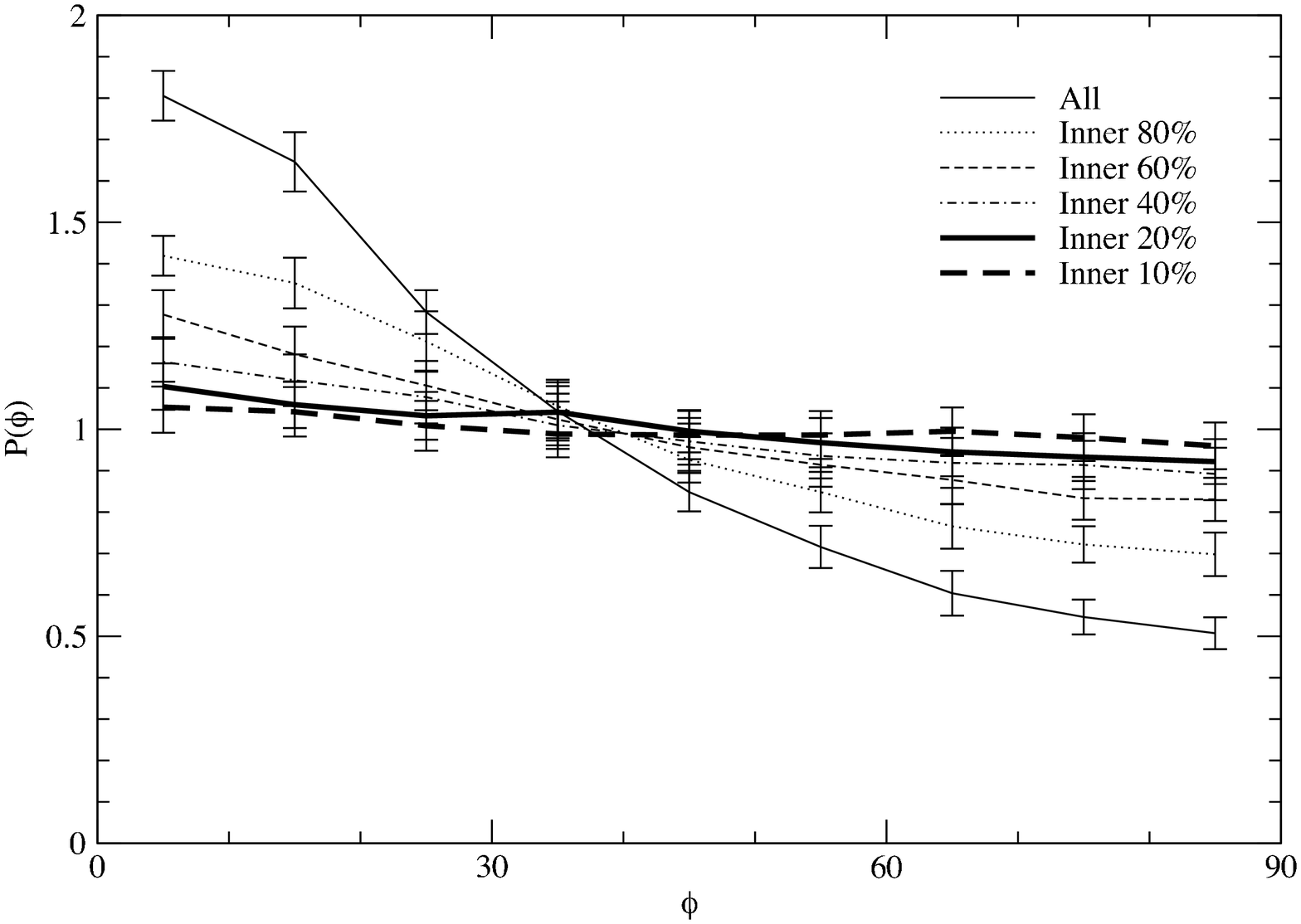, width=1\textwidth, angle=0}
\end{minipage}
\end{center}
\caption{Same as \Fig{fig:ra2Dmass}, but each line shows the correlation using major axes derived from a fraction of particles. The different lines represent shape measures at different
mass thresholds. Error bars indicate the projection variation.
\label{fig:ra2Drad}}
\end{figure}

As with many cosmological $N$-body studies, we are considering the
orientations of dark matter halos and are comparing them with the
distributions of baryons observed in galaxy clusters. Such a
comparison can be significantly biased if the observed orientation of
galaxies, which occupy the deepest part of the subhalo potential, are
not aligned with the global orientation of the dark matter. We
examined this effect by studying the dependence of the (projected)
radial alignment signal on the point where we measure the subhalos'
shape as outlined in \Sec{sec:s2Drad}.

The result of selecting the inner particles is quite striking;
\Fig{fig:ra2Drad} presents the orientation correlation using different
percentage of inner particles in subhalos. We observe a drop in the
signal's strength when restricting to the central parts; considering
only the inner 10-20\% of particles we recover the observed
correlation noted by~\citet{Pereira05}: following \citet{Struble85}
and \citet{Pereira05}, respectively, we quantify our results in terms
of an alignment parameter $\delta = \sum_i (\phi_i/N)-45^o$, where
$\phi_i$ are the indivudal subhalo orientations and $N$ the number of
subhalos. When using all particles in a subhalo we derive a value of
$\delta \approx -11^o$. However, when restricting to the innermost
20\% (10\%) we obtain $\delta \approx -2^o$ ($\delta \approx -1^o$) in
agreement with the value reported by \citet{Pereira05}. The weakening
of the signal could be due to poorly defined principle axes if the
inner distributions of particles were signficantly rounder than the
overall dark matter distribution.  However, we already noted that the
distribution of sphericity moves towards smaller values in the inner
parts of the subhalos (cf. \Fig{fig:s2Drad}). In that regards caution
must also be urged as, for the smallest halos containing of order 100
particles (cf. \Sec{sec:subhalos}), considering the inner 10\% of
particles means that we are measuring principle axes using of order 10
particles, and as such are very sensitive to Poisson noise.  In order
to avoid such small numbers of particles and to verify the credibility
of the weakening of the signal, we repeated the analysis restricting
the subhalos to those whose mass is in the range $10^{12}$\hMsun $\leq
M_{\mbox sub} < 10^{13}$\hMsun, that is the number of constituent
particles is in the range, $3289 \leq N_p < 32897$. We recover the
same orientation correlation, with the lower the fraction of inner
particles, the lower the observed signature of orientation
correlation, again bringing into agreement with the observed
correlation for the inner 10-20\% of particles.

\section{Summary and Conclusions}
\label{sec:summary}

We have examined whether or not the radial alignment of the major axes
of subhalos with respect to the centre of their host dark matter halo
is still present when projecting the data from cosmological
simulations into the observer's plane. Our results draw upon a sample
of dark matter host halos spanning group- to cluster-mass scales
($6\times 10^{13}$\hMsun\ to $6\times 10^{14}$\hMsun).

While observations of clusters of galaxies have revealed a mild radial
alignment of subhalos with respect to their host \citep{Pereira05,
  Wang07} the signal found in our data is substantially
stronger. However, when restricting the shape determination to using
only the innermost 10-20\% of the subhalo's particles we recover the
observed correlation strength, in agreement with the results reported by \citet{Faltenbacher07b}. We though note that the signal does not
weaken because the central parts of subhalos are rounder and hence
radial alignment itself becomes ill-defined. On the contrary, in
agreement with previous findings we observe the trend for subhalos to
become more aspherical when approaching the centre
\citep[cf. ][]{Jing02, Allgood06, Hayashi07}. Therefore, while the
strength of the radial alignment decreases even though the asphericity
increass towards the centre there has to be a misalignment between the
inner regions of the subhalo and its outer structure.  However, it is
known that the major axes of the central parts of field halos are
aligned with those of the whole object \citep[cf. ][]{Jing02}.  One
possible solution to this puzzling observation is tightly coupled to
the overall explanation of the radial alignment as a dynamical/tidal
effect \citep{Kuhlen07, Pereira07}: it can be surmised that subhalos
which have passed through their apocenter must experience a strong
tidal field and the outer -- more loosely bound -- regions of a
subhalo will be subject to more distortion than the inner parts.  This
results in a weaker correlation of the protected inner parts parts of
the sub-halo and the host.

We may even be as bold as to reverse the argumentation and use the
(projected) radial alignment to infer the point where the galaxy is
expected to lie within the dark matter subhalo: The point where the
strength of the signal matches the observed correlation -- that is
based upon the shape of the galaxy -- coincides with 10-20\% of the
total subhalo mass. 
Assuming a density profile of the functional form proposed by \citet[][NFW]{Navarro97} and a fiducial concentration of about $c$=10-15 for this profile, 15\% of $M_{\rm vir}$ is reached at about 9-11\% of the virial radius.\footnote{Even though tidal stripping will affect the outer regions of subhalos the central parts will still follow a (simple) power-law density profile as given by, for instance, the NFW profile \citep[cf. ][]{Kazantzidis04}.} This is in agreement with the results found by \citet{Bailin05} who found that this marks a point where the influence of the baryons becomes important and there appears to be a lack of connection between the inner and outer regions defined by this point, respectively.

\section*{Acknowledgements}
AK acknowledges funding through the Emmy Noether programme of the DFG
(KN 755/1). The simulations were
carried out on the super computer of the Center for
Computational Astrophysics, the National Astronomical Observatory of
Japan (Project-ID: why36b, ihy01b).  HY acknowledges the support of
the Research Fellowships of the Japan Society for the Promotion of
Science for Young Scientists (17-10511).  HY is also thankful to
Junichiro Makino and Toshiyuki Fukushige for their useful
comments. This research has been undertaken as part of the
Commonwealth Cosmology Initiative (\texttt{http://www.thecci.org}).

\bibliography{papers} \bsp

\label{lastpage}

\end{document}